\documentclass{moriond}

\def\be{\begin{equation}}
\def\ee{\end{equation}}
\def\bea{\begin{eqnarray}}
\def\eea{\end{eqnarray}}

\usepackage{amsmath}
\usepackage[font=small]{caption}

\newcommand{\Photo}{}

\begin{document}
\vspace*{4cm}
\title{Impact of Calibration Systematics on Dark Energy Constraints\\from LSST Type Ia Supernovae}

\author{Jonah Medoff$^{1}$, Christopher W. Stubbs$^{1}$, Dillon Brout$^{2}$ on behalf of the LSST-DESC collaboration}

\address{$^{1}$Department of Physics, Harvard University, 17 Oxford Street, Cambridge, MA, USA\\$^{2}$Department of Astronomy, Boston University, 725 Commonwealth Avenue, Boston, MA, USA}

\maketitle\abstracts{
The Vera C. Rubin Observatory’s Legacy Survey of Space and Time (LSST) will deliver an unprecedented Type Ia supernova (SN) sample, making photometric calibration systematics a dominant source of uncertainty in dark energy constraints. We perform a comprehensive analysis of calibration systematic effects in LSST, quantifying how uncertainties in the LSST passbands propagate into biases in SN distance moduli and, consequently, the dark energy equation of state parameters. Specifically, we examine how the inferred values and uncertainties of $w_0$ and $w_a$ shift as a function of the amplitude of passband systematics. For linear passband tilts, we find that the best-fit ($w_0$,$w_a$) shifts by $\sim$0.025$\sigma$ and the $w_0-w_a$ contour area increases by $\sim$5\% for each 1\%/100nm increase in tilt, while for quadratic passband tilts, our results are less conclusive and warrant further exploration. This analysis will help inform the calibration accuracy required for LSST to achieve its goals in constraining dark energy.}

\section{Introduction}

Rubin will observe hundreds of thousands of SNe Ia across the southern sky, drastically increasing the sample size available for cosmological analyses. With statistical uncertainties shrinking as 1/$\sqrt{\text{N}}$, systematic effects — particularly flux calibration — will dominate the error budget. In this regime, even millimagnitude-level calibration errors can impact the cosmological signal, making accurate passband calibration a central challenge for Rubin-era supernova cosmology. 

Photometric calibration uncertainties can arise from multiple sources, including imperfect knowledge of instrumental passbands and throughput, uncertainties in flux standards (e.g., CALSPEC white dwarfs), atmospheric transmission variability, and Galactic extinction corrections~\cite{StubbsBrown2016}. Importantly, recent analyses have shown that supernova miscalibration can introduce dataset tension and even produce spurious evidence for evolving dark energy~\cite{Dovekie,Ong2026}. These results underscore the need to quantify how calibration systematics propagate into cosmological inferences. 

We therefore assess the impact of LSST calibration uncertainties on dark energy constraints from SNIa measurements. Unlike previous LSST systematic analyses~\cite{Hazenberg2018,Mitra2025}, we focus specifically on LSST passband uncertainties and evaluate how shifts in cosmological parameters vary with systematic amplitude. Our primary aim is to determine how much effort needs to be put into calibrating Rubin such that it can achieve its dark energy science objectives~\cite{DESC_SRD}.

\section{Methods}

Instead of zeropoint or wavelength shifts, we distort the LSST passbands directly via linear and quadratic tilts. Perturbing the passband shape captures chromatic calibration errors that cannot be represented by simple zeropoint or wavelength offsets.
We intentionally introduce exaggerated passband tilts in order to clearly illustrate their impact on cosmological parameters. 

To perform this analysis, we use the supernova analysis software, \texttt{SNANA}~\cite{SNANA}. The general procedure is as follows: 1) We simulate an idealized (noise-free) sample of LSST Deep Drilling Field supernovae; 2) We inject calibration systematics by modifying the passbands in the K-Correction file used during light-curve fitting; 3) The resulting distorted passbands alter the inferred distance moduli, producing a shifted Hubble diagram and shifted cosmological parameters.

\section{Results}

\begin{figure}[h]
   \centering
   \begin{minipage}[c]{0.275\textwidth}
     \includegraphics[width=\linewidth]{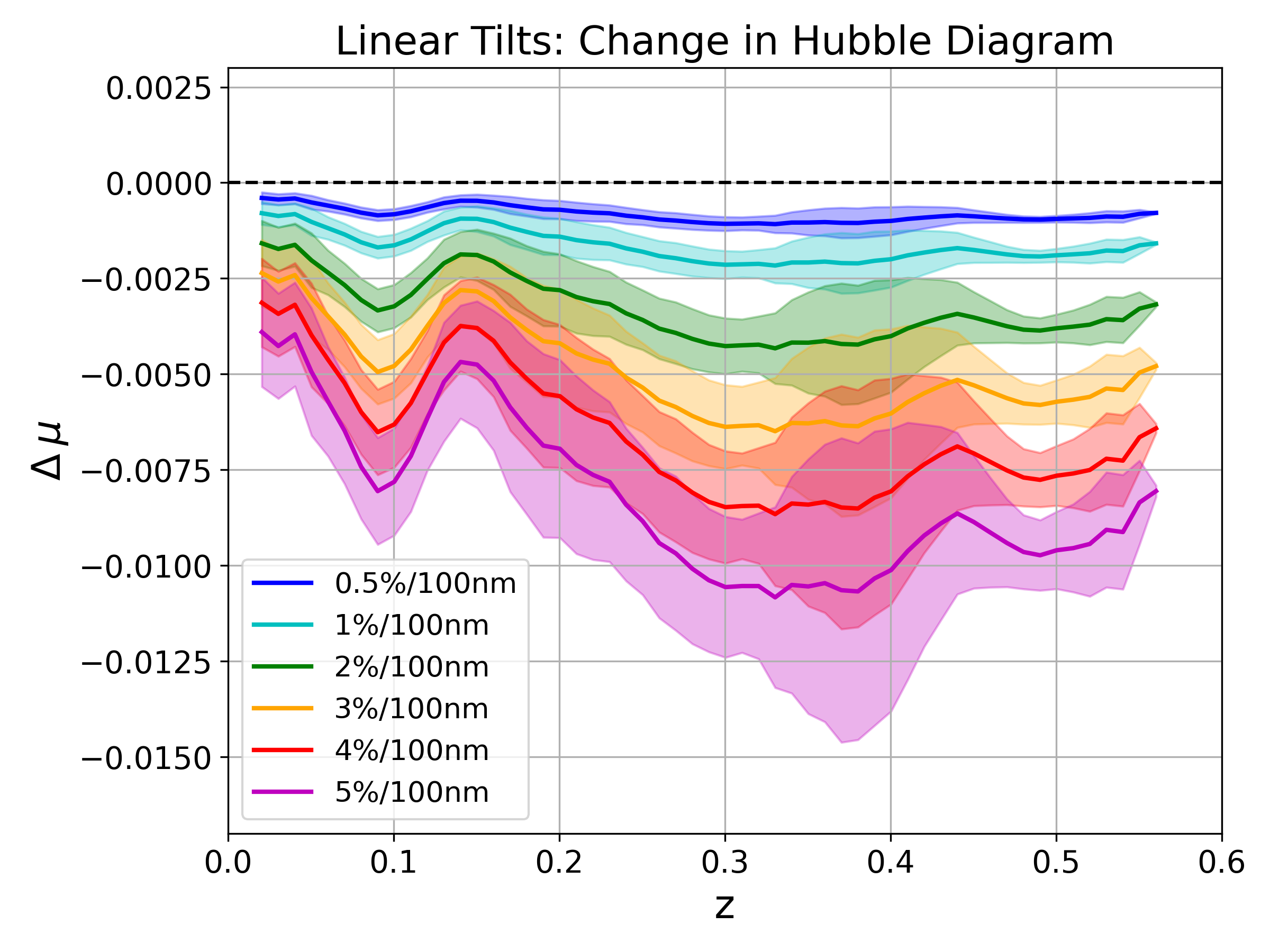}
     \vspace{2mm}
     \includegraphics[width=\linewidth]{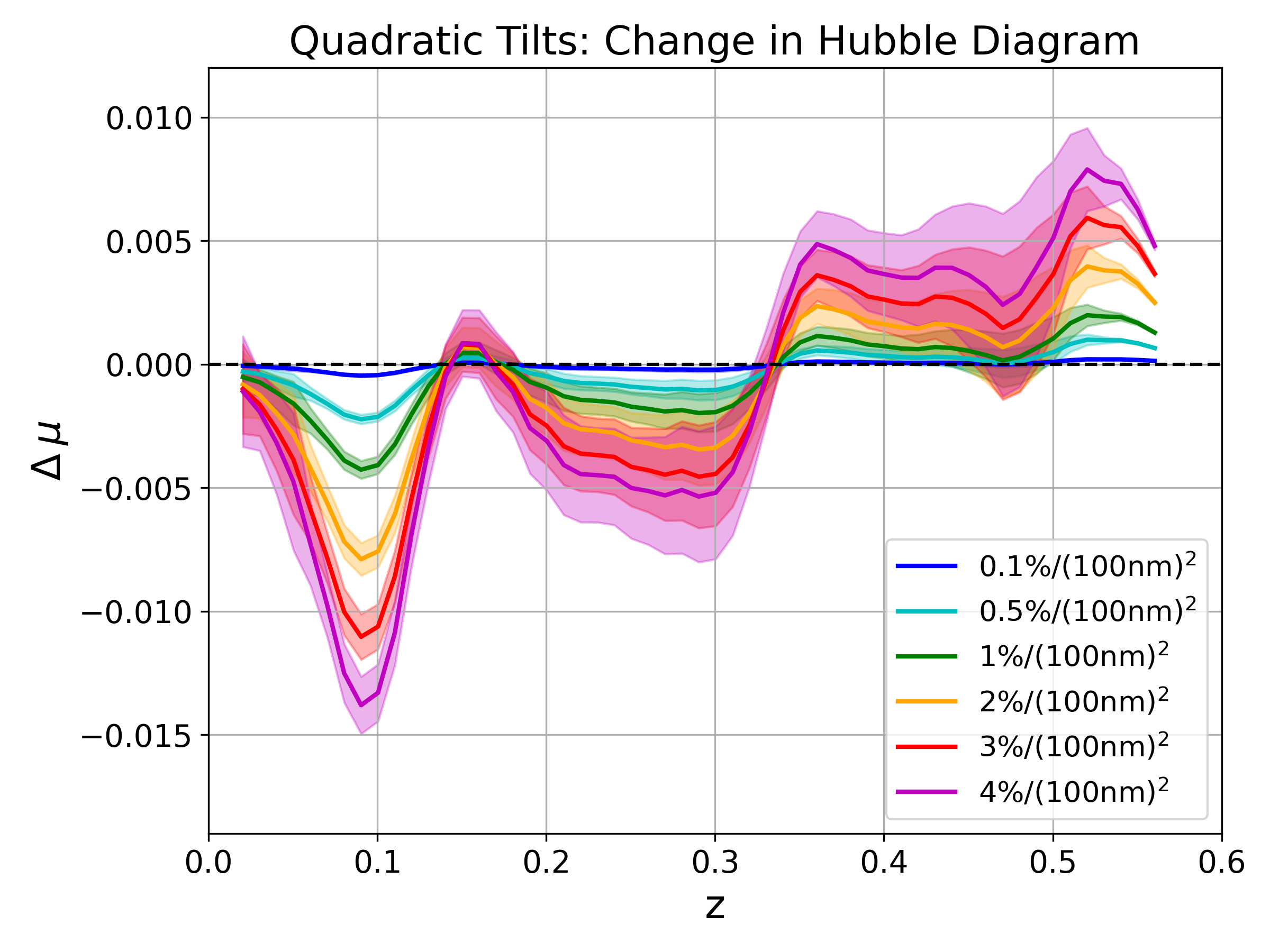}
     \caption{Changes in Hubble diagram from each linear (top) and quadratic (bottom) passband tilt. Mean $\Delta \mu$ given by solid lines. 1$\sigma$ (top) and 0.5$\sigma$ (bottom) error given by shaded regions.}
     \label{fig:left}
   \end{minipage}
   \hfill
   \raisebox{3mm}{
   \begin{minipage}[c]{0.43\textwidth}
     \includegraphics[width=\linewidth]{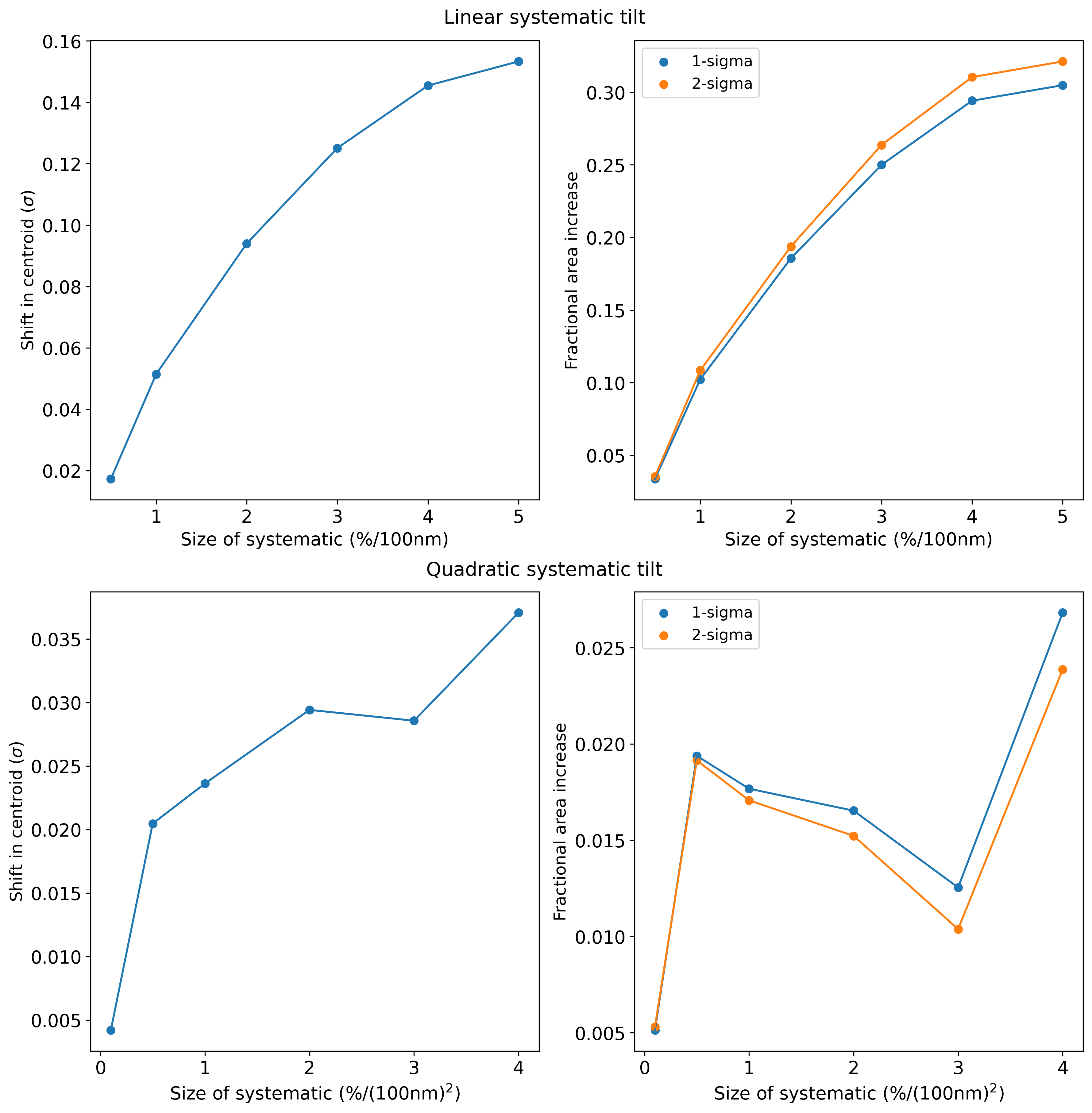}
     \caption{Changes in $w_0-w_a$ contours as a function of linear tilt coefficients (top row) and quadratic tilt coefficients (bottom row). Left:  Shift in contour centroid in units of $\sigma$. Right: Fractional increase in area of 1$\sigma$ (blue) and 2$\sigma$ (orange) contours.}
     \label{fig:middle}
   \end{minipage}
   }
   \hfill
   \raisebox{1.25mm}{
   \begin{minipage}[c]{0.245\textwidth}
     \includegraphics[width=\linewidth]{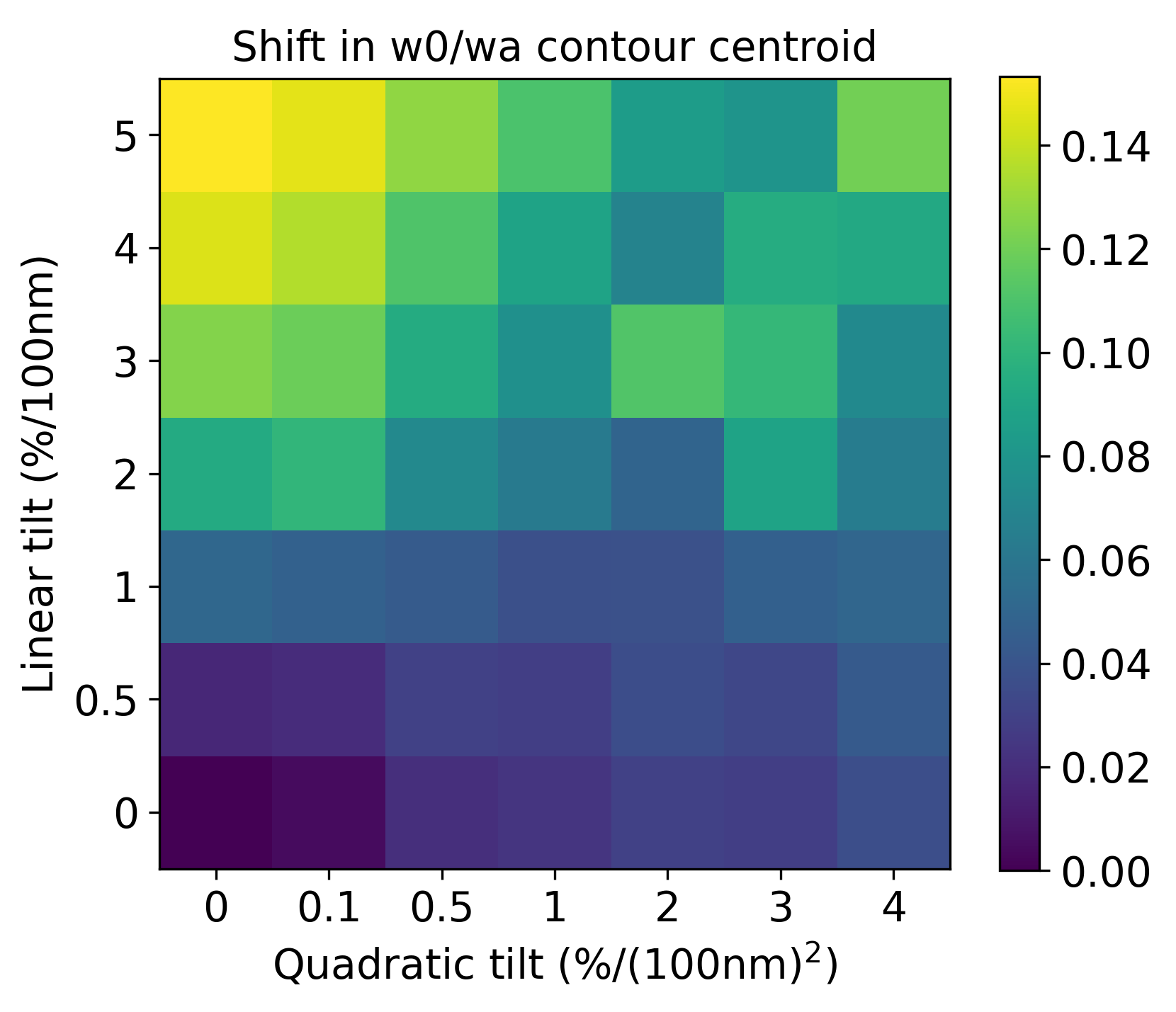}
     \vspace{2mm}
     \includegraphics[width=\linewidth]{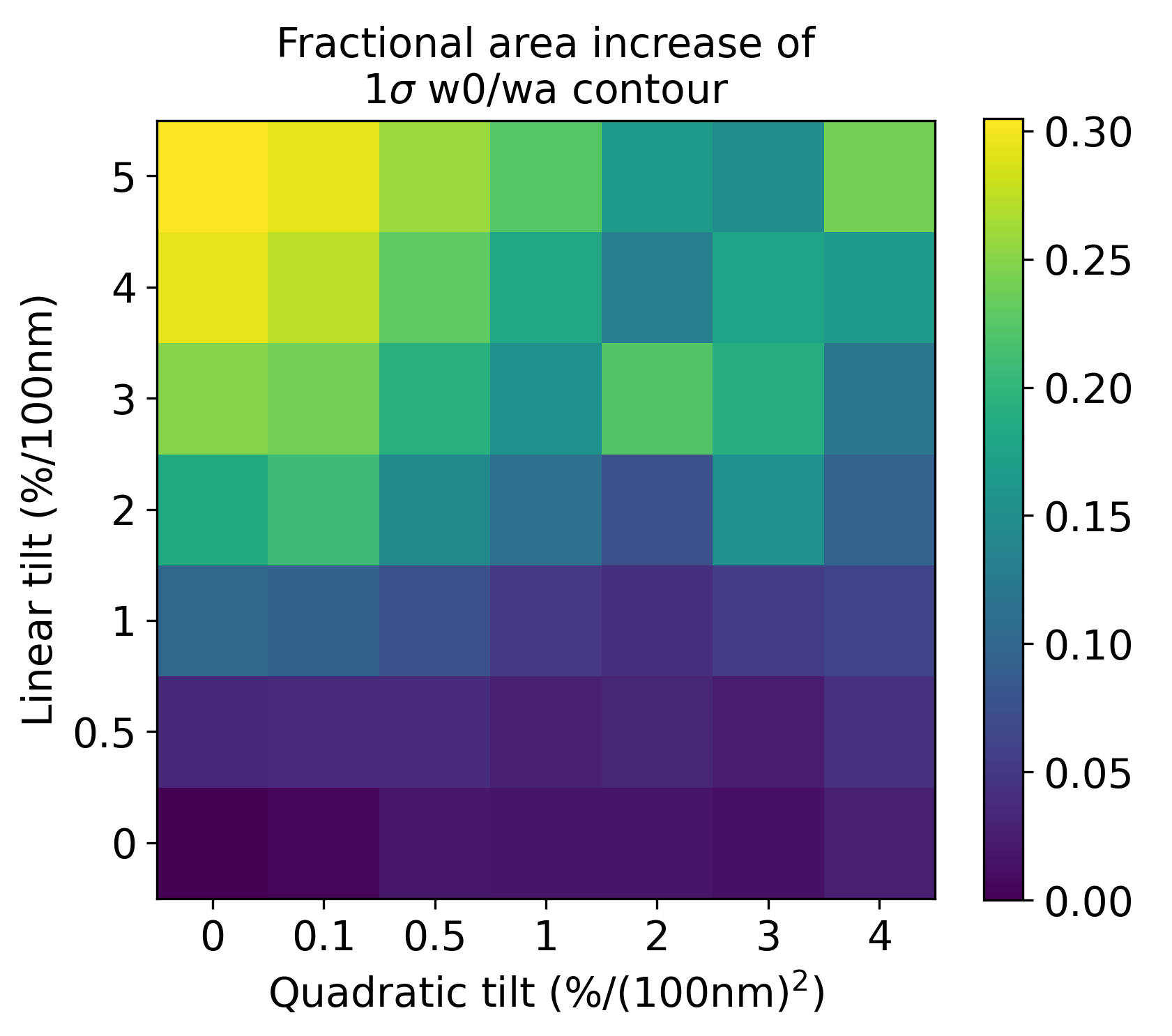}
     \caption{Top: Shift in $w_0-w_a$ contour centroid as a 2D function of linear and quadratic tilts. Bottom: Fractional area increase as a 2D function of linear and quadratic tilts.}
     \label{fig:right}
   \end{minipage}
   }
 \end{figure}

After repeating the above procedure for several linear and quadratic tilt coefficients, we show the resulting changes in the Hubble diagram (Figure \ref{fig:left}) and quantify the changes in the $w_0-w_a$ contours as a function of systematic size (Figure \ref{fig:middle}). We consider both the shift in contour centroid (best-fit values) and fractional increase in contour area. In addition to varying linear and quadratic tilts separately, we evaluate the contour changes over a 2D grid of linear and quadratic coefficients (Figure \ref{fig:right}).


\section*{Acknowledgments}

We gratefully acknowledge support from the US DOE under Cosmic Frontier grant DESC0007881.

\section*{References}
\bibliography{moriond}


\end{document}